\documentclass[twocolumn,preprintnumbers,amsmath,amssymb]{revtex4}

\usepackage{graphicx}
\usepackage{dcolumn}
\usepackage{bm}

\begin{document}


\title{Sagnac interferometry and non--Newtonian gravity}
\author{Abel Camacho}
\email{acq@xanum.uam.mx}
\affiliation{Department of Physics \\
Universidad Aut\'onoma Metropolitana--Iztapalapa\\
Apartado Postal 55--534, C.P. 09340, M\'exico, D.F., M\'exico.}

\author{ A. Camacho--Galv\'an}
\email{abel@servidor.unam.mx}
\affiliation{ DEP--Facultad de Ingenier{\'\i}a\\
Universidad Nacional Aut\'onoma de M\'exico.}

\date{\today}

\begin{abstract}
Sagnac interferometry has been employed in the context of gravity
as a proposal for the detection of the so called gravitomagnetic
effect. In the present work we explore the possibilities that this
experimental device could open up in the realm of non--Newtonian
gravity. It will be shown that this experimental approach allows
us to explore an interval of values of the range of the new force
that up to now remains unexplored, namely, $\lambda\geq 10^{14}$
m.

\end{abstract}
\maketitle

\section{Introduction}

The use of interferometric techniques has rendered very fruitful
results in the realm of gravitation, as a fleeting glimpse to the
experimental efforts in gravitational waves \cite{[1]} or in the
detection of the so called gravitomagnetic effect \cite{[2]},
easily attests. We may rephrase this stating that optical
interferometry has played along many years a fundamental role in
gravitational physics.

Though general relativity is one of the bedrocks of modern
physics, and many of its predictions have found a sound
confirmation at the experimental level \cite{[3]}, the quest for
deviations from the predictions of Newtonian gravity has never
waned altogether, since many theoretical attempts to construct a
model of elementary particles do predict the emergence of new
forces (usually denoted as fifth force) \cite{[4]}. One of the
distinctive traits of these new interactions is the fact that they
are not described by a inverse--square law. Additionally, they,
generally, violate the so called weak equivalence principle
\cite{[5]}. Since more than a decade has witnessed the lack of any
kind of compelling evidence that could purport some kind of
deviations from the Newtonian theory \cite{[6]}, the pursuance in
this direction requires a thorough justification, a requirement
already covered by Gibbons and Whiting \cite{[7]}, who claim that
a very precise agreement between Newtonian gravity and the
observation of planetary motion does not preclude the existence of
large non--Newtonian effects over other distance scales. The
results comprised in \cite{[6]} allow us to draw several
conclusions: (i) the cu\-rrent experimental constraints do not
explore the so called geophysical window ($\lambda\in [10m,
1000m]$); (ii) the case in which $\lambda\geq 10^{14}$ m remains
completely unexplored.

The main goal in the present work is the introduction of an
experimental proposal, which could explore the region $\lambda\geq
10^{14}$ m. This will be done by means of an experimental model
embracing a Sagnac interferometer \cite{[8]} whose area (that
enclosed by the light path) has a unit normal vector perpendicular
to the direction of the acceleration of gravity. It will be shown
that this idea may be used to find the first experimental bound in
the aforementioned region of $\lambda$.

\section{Sagnac Interferometry and non--Newtonian gravity}

Let us consider a gravitational potential which contains a
Yukawa--type term \cite{[9]}

\begin{equation}
U(r) = - \frac{G_{\infty}M}{r}\Bigl\{1 +
\alpha\exp\{-r/\lambda\}\Bigr\}. \label{Gravpot}
\end{equation}

Here $G_{\infty}$ denotes the value of the Newtonian constant
between the source of the gravitational field, i.e., $M$, and a
test particle when the distance between them tends to infinity. As
a matter of fact $G_N = G_{\infty}(1 + \alpha)$, where $G_N$ is
the usual Newtonian constant. In addition, $\lambda$ is the range
of the interaction.

At this point, and bearing in mind that we try to put forward a
terrestrial experimental proposal, the following approximation
will be introduced, $r = R + z$, with $R>>\vert z\vert$. Under
this restriction (\ref{Gravpot}) becomes, to first order in $z/R$

\begin{equation}
U(r) = - \frac{G_{\infty}M}{r}\Bigl\{1 - \frac{z}{R} +
\alpha\exp\{-R/\lambda\}\Bigl(1 -  \frac{R
+\lambda}{\lambda}\frac{z}{R}\Bigr)\Bigr\}. \label{Gravlim}
\end{equation}

At this point consider now a Sagnac interferometer whose area
(that enclosed by the light path) has a normal vector
perpendicular to the direction of the acceleration of gravity,
i.e., the $z$--axis. In addition, the angular velo\-city of the
interferometer (the one rotates in the clockwise direction) is
$\Omega$, and its radius $a$. For the sake of simplicity let us
assume that the beams enter the interferometer at point $A$, which
is the highest one, (its $z$ is a maximum). Since the
interferometer rotates, then both beams meet, for the first time,
at \cite{[8]}

\begin{equation}
t_d = \frac{2\pi a}{c}\Bigl\{1 + \frac{a\Omega}{c}\Bigr\}^{-1}.
\label{Time1}
\end{equation}

The distance, below point $A$, at which the beams meet for the
first time is

\begin{equation}
h = a\Bigl\{1 - \cos\Bigl(\frac{2\pi\Omega a}{c}[1 +
\frac{a\Omega}{c}]^{-1}\Bigr)\Bigr\}.
 \label{Height}
\end{equation}

Since the beams are immersed in a region in which the
gravitational potential has the form pointed out in
(\ref{Gravlim}), then during their movement they will undergo a
red--shift, the one reads \cite{[3]}

\begin{equation}
\nu_{rs} = \frac{\nu}{1 + \Delta U/c}.
 \label{Red}
\end{equation}

Here $\Delta U$ denotes the difference in the potential between
the two involved points.

The frequency at time $t_d$ reads

\begin{equation}
\nu_{rs} = \frac{\nu}{1 - a\frac{G_{\infty}M}{c^2R^2}}\gamma\beta.
\label{Freq}
\end{equation}

In this last expression two definitions have been introduced

\begin{equation}
\beta = \Bigl\{1 + \alpha\frac{R +
\lambda}{\lambda}\exp\{-R/\lambda\}\Bigr\}^{-1}, \label{Defin1}
\end{equation}

\begin{equation}
\gamma = \Bigl\{1 - \cos\Bigl(\frac{2\pi\Omega a}{c}[1 +
\frac{a\Omega}{c}]^{-1} \Bigr)\Bigr\}^{-1}. \label{Defin2}
\end{equation}

The time difference between the arrival of the two beams is the
usual one \cite{[8]}

\begin{equation}
\Delta t = \frac{4\pi a^2\Omega}{c^2 - a^2\Omega^2}.
 \label{Timediff}
\end{equation}

This last result renders the path difference, $\Delta L = c\Delta
t$.

\begin{equation}
\Delta L = \frac{4\pi a^2c\Omega^2}{c^2 - a^2\Omega^2}.
\label{Pathdiff}
\end{equation}

Finally, harking back to (\ref{Freq}) the phase difference,
$\Delta\theta$ reads

\begin{equation}
\Delta\theta = \frac{8\pi^2 a^2\nu\Omega}{(c^2 - a^2\Omega^2)(1 -
a\frac{G_{\infty}M}{c^2R^2})}\gamma\beta. \label{Phasediff}
\end{equation}

Writing this phase difference as the sum of two terms,
$\Delta\theta^{(N)}$ and $\Delta\theta^{(NN)}$, which correspond
to the differences stemming from the Newtonian and non--Newtonian
parts of the gravitational potential, respectively, we may deduce
(assuming $\vert\alpha\frac{R
+\lambda}{\lambda}\exp\{-R/\lambda\}\vert <1$)

\begin{equation}
\Delta\theta^{(NN)} = -\Delta\theta^{(N)}\alpha\frac{R +
\lambda}{\lambda}\exp\{-R/\lambda\}.\label{Nnphasediff}
\end{equation}

In this result we have included the fact that

\begin{equation}
\Delta\theta^{(N)} = \frac{8\pi^2 a^2\nu\Omega}{(c^2 -
a^2\Omega^2)(1 - a\frac{G_{\infty}M}{c^2R^2})}\gamma. \label{Def}
\end{equation}

\section{Conclusions}

The possible detection of a fifth force through this kind of
proposals does strongly depend upon the relation between the
experimental resolution associated to the measuring process of
phase differences, $\Delta\theta^{(ex)}$, and the absolute value
of the parameter $\Delta\theta^{(NN)}/\Delta\theta^{(N)}$. In
other words, this idea could be a useful one if

\begin{equation}
\vert\Delta\theta^{(NN)}/\Delta\theta^{(N)}\vert >
\Delta\theta^{(ex)}. \label{Det}
\end{equation}

Resorting to (\ref{Nnphasediff}) and (\ref{Def}) it is readily
seen that the feasibility of the proposal becomes

\begin{equation}
\vert\alpha\vert\frac{R + \lambda}{\lambda}\exp\{-R/\lambda\} >
\Delta\theta^{(ex)}. \label{Det1}
\end{equation}

A fleeting glimpse at the current experimental bounds \cite{[6]}
immediately shows us that several regions of $\lambda$ remain
unexplored. For instance, there are no experiments related to
$\lambda \geq 10^{14}$ m. Bearing in mind that we should
contemplate the possibility of performing this experiment on the
Earth's surface, then $R\sim 10^{6}$m, and if we consider the
aforementioned region for the range of the fifth force, then
$R/\lambda\sim 10^{-8}$, and hence

\begin{equation}
\vert\alpha\vert > \Delta\theta^{(ex)}. \label{Det2}
\end{equation}

We may rephrase this last conclusion asserting that the resolution
of the measuring device sets (in this very particular situation)
the bound upon the values of the strength . If nothing is seen,
then, for sure

\begin{equation}
\vert\alpha\vert < \Delta\theta^{(ex)}. \label{Det3}
\end{equation}

Let us now address the issue concerning the order of magnitude of
$\alpha$ that could be measured in this kind of models. It is
already known that the resolving power of a spectral device is
defined as $\Lambda = \tilde{\lambda}/\Delta\tilde{\lambda}$
\cite{[10]}, here $\tilde{\lambda}$ denotes the wavelength of the
corresponding beam, while $\Delta\tilde{\lambda}$ is the minimum
difference in the wavelength which can be resolved. If the
experiment is carried out by means of visible light
$(\tilde{\lambda}\sim 10^{-6}$m), and assuming a rough resolving
power, i.e., $\Lambda\sim 10^{5}$, then (since
$\Delta\theta^{(ex)}\sim 2\pi/\Lambda$) we obtain that
$\Delta\theta^{(ex)}\sim 10^{-4}$. Under these constraints it
becomes clear that a null experiment would imply that
$\vert\alpha\vert\leq 10^{-4}$.

Summing up, it has been shown that a Sagnac interferometer can be
employed to impose an experimental bound for the strength of an
hypothetical fifth force, for the case, which up to now remains
completely unexplored, $\lambda\geq 10^{14}$m.

\begin{acknowledgments}
This research was supported by CONACYT Grant 42191--F. A.C. would
like to thank A.A. Cuevas--Sosa for useful discussions and
literature hints. \end{acknowledgments}

\end{document}